\documentclass[aps,prl,preprint,nopacs,superscriptaddress]{revtex4}

\usepackage{graphicx}
\usepackage{verbatim}
\usepackage{mathrsfs}
\usepackage{amsmath,amsfonts,amssymb}
\usepackage{graphicx}

\def\3{2.8in}    
\def\2{2.5in}
\def\4{3.0in}

\def \beq {\begin{equation}}
\def \eeq {\end{equation}}
\begin{document}

\title{Arc-tunable Weyl Fermion metallic state in Mo$_{x}$W$_{1-x}$Te$_2$}
\author{Tay-Rong Chang}
\affiliation{Department of Physics, National Tsing Hua University, Hsinchu 30013, Taiwan}

\author{Su-Yang Xu}\affiliation {Laboratory for Topological Quantum Matter and Spectroscopy (B7), Department of Physics, Princeton University, Princeton, New Jersey 08544, USA}

 \author{Guoqing Chang}
 \affiliation{Centre for Advanced 2D Materials and Graphene Research Centre National University of Singapore, 6 Science Drive 2, Singapore 117546}
 \affiliation{Department of Physics, National University of Singapore, 2 Science Drive 3, Singapore 117542}
 
 \author{Chi-Cheng Lee}
 \affiliation{Centre for Advanced 2D Materials and Graphene Research Centre National University of Singapore, 6 Science Drive 2, Singapore 117546}
 \affiliation{Department of Physics, National University of Singapore, 2 Science Drive 3, Singapore 117542}
 
 \author{Shin-Ming Huang}
 \affiliation{Centre for Advanced 2D Materials and Graphene Research Centre National University of Singapore, 6 Science Drive 2, Singapore 117546}
 \affiliation{Department of Physics, National University of Singapore, 2 Science Drive 3, Singapore 117542}
 
 \author{BaoKai Wang}
 \affiliation{Centre for Advanced 2D Materials and Graphene Research Centre National University of Singapore, 6 Science Drive 2, Singapore 117546}
 \affiliation{Department of Physics, National University of Singapore, 2 Science Drive 3, Singapore 117542}
 \affiliation{Department of Physics, Northeastern University, Boston, Massachusetts 02115, USA}
 
\author{Guang Bian}\affiliation {Laboratory for Topological Quantum Matter and Spectroscopy (B7), Department of Physics, Princeton University, Princeton, New Jersey 08544, USA}
\author{Hao Zheng}\affiliation {Laboratory for Topological Quantum Matter and Spectroscopy (B7), Department of Physics, Princeton University, Princeton, New Jersey 08544, USA}
\author{Daniel S. Sanchez}\affiliation {Laboratory for Topological Quantum Matter and Spectroscopy (B7), Department of Physics, Princeton University, Princeton, New Jersey 08544, USA}
\author{Ilya Belopolski}\affiliation {Laboratory for Topological Quantum Matter and Spectroscopy (B7), Department of Physics, Princeton University, Princeton, New Jersey 08544, USA}
\author{Nasser Alidoust}\affiliation {Laboratory for Topological Quantum Matter and Spectroscopy (B7), Department of Physics, Princeton University, Princeton, New Jersey 08544, USA}
\author{Madhab Neupane}\affiliation {Laboratory for Topological Quantum Matter and Spectroscopy (B7), Department of Physics, Princeton University, Princeton, New Jersey 08544, USA}
\affiliation {Condensed Matter and Magnet Science Group, Los Alamos National Laboratory, Los Alamos, NM 87545, USA}

 \author{Arun Bansil}
 \affiliation{Department of Physics, Northeastern University, Boston, Massachusetts 02115, USA}
 \author{Horng-Tay Jeng}
 \affiliation{Department of Physics, National Tsing Hua University, Hsinchu 30013, Taiwan}
 \affiliation{Institute of Physics, Academia Sinica, Taipei 11529, Taiwan}
 
 \author{Hsin Lin*}
 \affiliation{Centre for Advanced 2D Materials and Graphene Research Centre National University of Singapore, 6 Science Drive 2, Singapore 117546}
 \affiliation{Department of Physics, National University of Singapore, 2 Science Drive 3, Singapore 117542}
 
\author{M. Zahid Hasan\footnote{Corresponding authors (emails): nilnish@gmail.com and mzhasan@princeton.edu}}\affiliation {Laboratory for Topological Quantum Matter and Spectroscopy (B7), Department of Physics, Princeton University, Princeton, New Jersey 08544, USA}
\affiliation{Princeton Center for Complex Materials, Princeton Institute for the Science and Technology of Materials, Princeton University, Princeton, New Jersey 08544, USA}

\date{\today}

\begin{abstract}
Weyl semimetals may open a new era in condensed matter physics because they provide the first example of Weyl fermions, realize a new topological classification even though the system is gapless, exhibit Fermi arc surface states and demonstrate the chiral anomaly and other exotic quantum phenomena. So far, the only known Weyl semimetals are the TaAs class of materials. Here, we propose the existence of a tunable Weyl metallic state in Mo$_{x}$W$_{1-x}$Te$_2$ via our first-principles calculations. We demonstrate that a 2\% Mo doping is sufficient to stabilize the Weyl metal state not only at low temperatures but also at room temperatures. We show that, within a moderate doping regime, the momentum space distance between the Weyl nodes and hence the length of the Fermi arcs can be continuously tuned from zero to $\sim3\%$ of the Brillouin zone size via changing Mo concentration, thus increasing the topological strength of the system. Our results provide an experimentally feasible route to realizing Weyl 
physics in the layered compound Mo$_{x}$W$_{1-x}$Te$_2$, where non-saturating magneto-resistance and pressure driven superconductivity have been observed.

\end{abstract}

\maketitle


In 1929, H. Weyl noted that the Dirac equation takes a simple form if the mass term is set to zero \cite{Weyl}: $i\partial_{L}\Psi = c\vec{p}\cdot\vec{\sigma}\Psi$ , with $\sigma^{\alpha}$ being the conventional Pauli matrices. Such a particle, the Weyl fermion, is massless but is associated with a definite chirality. Weyl fermions may be thought of as the basic building blocks for a Dirac fermion. They have played a vital role in quantum field theory but they have not been found as fundamental particles in vacuum. A Weyl semimetal is a solid state crystal that host Weyl fermions as its low energy quasiparticle excitations \cite{Weyl, Herring, Abrikosov, Volovik2003, Murakami2007, Wan2011, Ran, Balents_viewpoint, Burkov2011,Hosur,Ojanen,Ashvin,Hasan2010,Qi2011,TI_book_2014, Hasan_Na3Bi}. Weyl semimetals have attracted intense research interest not only because they provide the only known example of a Weyl fermion in nature, but also because they can be characterized by a set of topological invariants even 
though the system is not an (topological) insulator. In a Weyl semimetal, a Weyl fermion is associated with an accidental degeneracy of the band structure. Away from the degeneracy point, the bands disperse linearly and the spin texture is chiral, giving rise to a quasiparticle with a two-component wavefunction, a fixed chirality and a massless, linear dispersion. Weyl fermions have distinct chiralities, either left-handed or right-handed. In a Weyl semimetal crystal, the chiralities of the Weyl nodes gives rise to topological charges, which can be understood as monopoles and anti-monopoles of Berry flux in momentum space. Remarkably, the topological charges in a Weyl semimetal are protected only by the translational invariance of the crystal. The band structure degeneracies in Weyl semimetals are uniquely robust against disorder, in contrast to the Dirac nodes in graphene, topological insulator and Dirac semimetals which depend on additional symmetries beyond the translational symmetry \cite{Hasan2010,
Qi2011, Hasan_Na3Bi, TI_book_2014, Graphene}. As a result, the Weyl fermion carriers are expected to transmit electrical currents effectively. Moreover, the transport properties of Weyl semimetals are predicted to show many exotic phenomena including the negative magnetoresistance due to the chiral anomaly known from quantum field theory, non-local transport and quantum oscillations where electrons move in real space between opposite sides of a sample surface \cite{Hosur,Ojanen,Ashvin}. These novel properties suggest Weyl semimetals as a flourishing field of fundamental physics and future technology. The separation of the opposite topological charges in momentum space leads to surface state Fermi arcs which form an anomalous band structure consisting of open curves that connect the projections of opposite topological charges on the boundary of a bulk sample. Without breaking symmetries, the only way to destroy the topological Weyl phase is to annihilate Weyl nodes with opposite charges by bringing them 
together in $k$ space. Thus the length of the Fermi arc provides a measure of the ``topological strength'' of a Weyl state.

For many years, research on Weyl semimetals has been held back due to the lack of experimentally feasible candidate materials. Recently, it was proposed that a family of isostructural compounds, TaAs, NbAs, TaP and NbP, are Weyl semimetals \cite{Huang2015, Hasan_TaAs, NbAs_Hasan, Weng2015}. Shortly after the theoretical prediction, the first Weyl semimetal was experimentally discovered in TaAs \cite{Hasan_TaAs}. So far, the TaAs class of four iso-electronic compounds remains to be only experimentally realized Weyl semimetals \cite{Hasan_TaAs,TaAs_Ding,TaAs_Ding_2,NbAs_Hasan}.

Tungsten ditelluride, WTe$_2$, has an inversion symmetry breaking crystal structure, and exhibits a compensated semi-metallic ground state \cite{WT-tran-1, WT-ARPES-1, WT-ARPES-2, WT-ARPES-3}. The coexistence of inversion symmetry breaking and semimetallic transport behavior resembles the properties of TaAs and hence suggests a possible Weyl semimetal state. Here, we propose a tunable Weyl metallic state in Mo-doped WTe$_2$ via our first-principles calculation, where the length of the Fermi arc and hence the topological strength of the system can be adiabatically tuned as a function of Mo doping. A very recent paper \cite{WT-Weyl} predicted the Weyl state in pure WTe$_2$ but the separation between Weyl nodes was reported to be beyond spectroscopic experimental resolution. We demonstrate that a 2\% Mo doping is sufficient to stabilize the Weyl metal state not only at low temperatures but also at room temperatures. We show that, within a moderate doping regime, the momentum space distance between the Weyl 
nodes and hence the length of the Fermi arcs can be continuously tuned from zero to $\sim3\%$ of the Brillouin zone size via changing Mo concentration, thus increasing the topological strength of the system. Our results present a tunable topological Weyl system, which is not known to be possible in the TaAs class of Weyl semimetals.

We computed the electronic structures using the projector augmented wave method \cite{PAW-1,PAW-2} as implemented in the VASP package \cite{VASP} within the generalized gradient approximation (GGA) schemes \cite{Perdew}. For WTe$_2$, experimental lattice constants were used \cite{WT-Structure}. For MoTe$_2$, we assumed that it has the same crystal structure as WTe$_2$ and calculated the lattice constants self-consistently ($a=6.328 \textrm{\AA}, b=3.453 \textrm{\AA}, c=13.506 \textrm{\AA}$). A $8\times16\times4$ MonkhorstPack $k$-point mesh was used in the computations. The spin-orbit coupling effects were included in calculations. In order to systematically calculate the surface and bulk electronic structure, we constructed a first-principles tight-binding model Hamilton for both WTe$_2$ and MoTe$_2$, where the tight-binding model matrix elements were calculated by projecting onto the Wannier orbitals \cite{wan-1,wan-2,wan-3}, which used the VASP2WANNIER90 interface \cite{Franchini}. The electronic 
structure of the samples with finite dopings was calculated by a 
linear interpolation of tight-binding model matrix elements of WTe$_2$ and MoTe$_2$. The surface state electronic structure was calculated by the surface Green's function technique, which computes the spectral weight near the surface of a semi-infinite system. We used W (Mo) $s$ and $d$ orbitals and Te $p$ orbitals to construct Wannier functions without using the maximizing localization procedure. 

\begin{figure*}
\includegraphics[width=16cm]{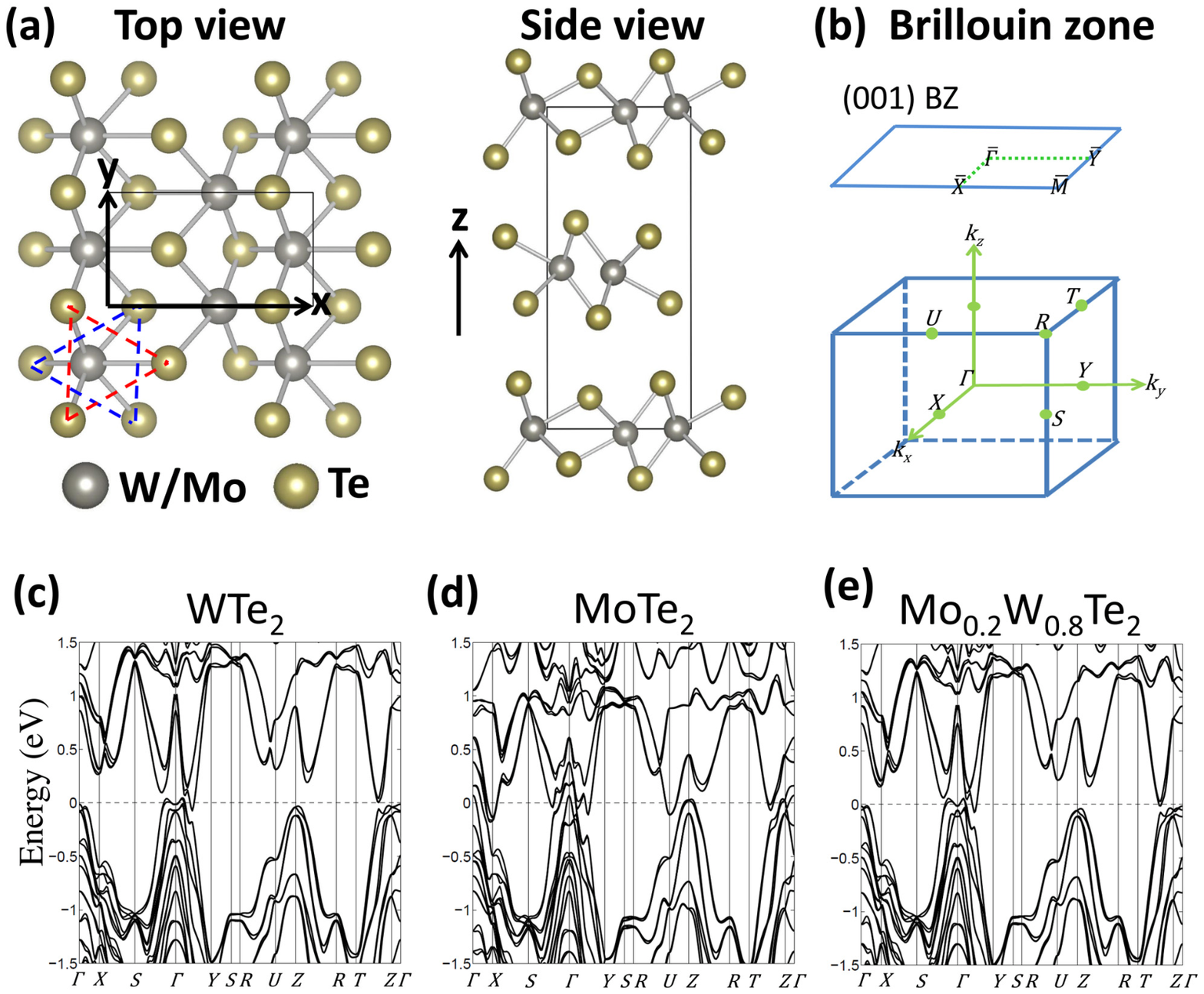}
\caption{ 
{\bf Crystal structure and band structure.} {\bf a,} A top view and a side view of WTe$_2$ lattice. The silver and yellow balls represent W and Te atoms, respectively. WTe$_2$ crystalizes in an orthorhombic Bravais lattice, space group $Pmn2_1$ (31). The lattice constants are $a=6.282$ $\textrm{\AA}$, $b=3.496$ $\textrm{\AA}$, $c=14.07$ $\textrm{\AA}$ according to a previous X-ray diffraction (XRD) measurement \cite{WT-Structure}. {\bf b,} The bulk and (001) surface Brillouin zone of WTe$_2$. {\bf c,} Bulk band structure of WTe$_2$. {\bf d,} Bulk band structure of MoTe$_2$ by assuming that it has the same crystal structure as WTe$2$. Note that, in fact, according to available literature MoTe$_2$ has two possible structures \cite{WT-Structure, MT-Structure}, both of which are different from the crystal structure of WTe$_2$. {\bf e,} Bulk band structure of Mo$_{0.2}$W$_{0.8}$Te$_{2}$.}
\label{band}
\end{figure*}

WTe$_2$ crystalizes in an orthorhombic Bravais lattice, space group $Pmn2_1$ (31). In this structure, each tungsten layer is sandwiched by two tellurium layers and form strong ionic bonds. The left panel of Fig.~\ref{band}(a) shows a top view of the lattice. It can be seen that the tungsten atom is shifted away from the center of the hexagon formed by the tellurium atoms. This makes the in-plane lattice constant along the $\hat{x}$ direction ($a$) longer than that of along the $\hat{y}$ direction ($b$). The WTe$_2$ sandwich stacks along the out of plane $\hat{z}$ direction, with van der Waals bonding between layers (the right panel of Fig.~\ref{band}(a)). We used the experimental lattice constants reported in Ref. \cite{WT-Structure}, $a=6.282$ $\textrm{\AA}$, $b=3.496$ $\textrm{\AA}$, $c=14.07$ $\textrm{\AA}$. The bulk Brillouin zone (BZ) and the (001) surface BZ are shown in Fig.~\ref{band}(b), where high symmetry points are noted. In Fig.~\ref{band}(c), we show the bulk band structure of WTe$_2$ along 
important high symmetry directions. Our calculation shows that there is a continuous energy gap near the Fermi level, but the conduction and valence bands have a finite overlap in energy. The band gap along the $\Gamma-Y$ direction is much smaller than that of along the $\Gamma-X$ direction or $\Gamma-Z$ direction, consistent with the fact that the lattice constant $b$ is much smaller than $a$ and $c$. At the Fermi level, our calculation reveals a hole pocket and an electron pocket along the $\Gamma-Y$ direction (in Fig.~\ref{band}(c)), which agrees with previous calculation and photoemission results \cite{WT-tran-1, WT-ARPES-1, WT-ARPES-2, WT-ARPES-3}. We also calculated the band structure of MoTe$_2$ by assuming that it is in the same crystal structure. As shown in Fig.~\ref{band}(d), the general trend is that the bands are ``pushed'' closer to the Fermi level. For example, in MoTe$_2$, there are bands crossing the Fermi level even along the $\Gamma-X$ and $\Gamma-Z$ directions. We emphasize that, 
according to available literature \cite{WT-Structure, MT-Structure}, MoTe$_2$ has a different crystal structure, either hexagonal \cite{MT-Structure} or monoclinic \cite{WT-Structure} both of which have inversion symmetry, but not orthorhombic. Thus in our calculation we assumed that MoTe$_2$ has the orthorhombic crystal structure as WTe$_2$ and obtained the lattice constants and atomic coordinates from first-principle calculations. Very recently, a paper \cite{MT-SC} claimed that MoTe$_2$ can be grown in the orthorhombic structure. This still needs to be further confirmed.  

\begin{figure*}[t]
\includegraphics[width=17cm]{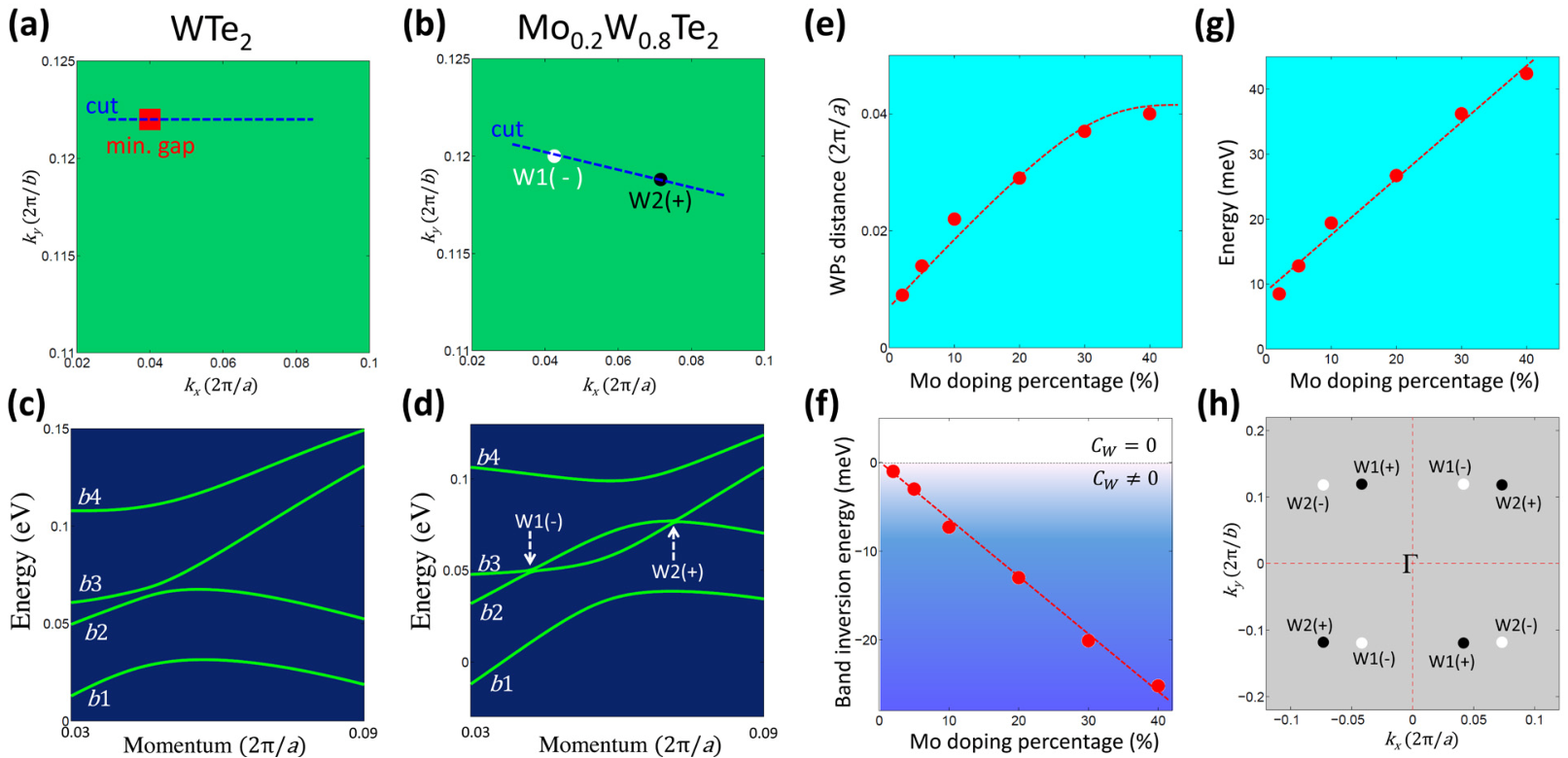}
\caption{ 
{\bf Tunable Weyl node separation (topological strength) in Mo-doped WTe$_2$.} {\bf a,} The $k$ space location that corresponds to the minimal energy gap of the pure (undoped) WTe$_2$. {\bf b,} The $k$ space locations of the pair of Weyl nodes within the first quadrant of the $k_z=0$ plane for Mo$_{0.2}$W$_{0.8}$Te$_2$. White and Black dots show the negative and positive chiral charges ($C_w$), respectively. {\bf c,} Band structures of WTe$_2$ along momentum space cut as defined in pane (a). The four bands appeared in this panel are labeled as b1-b4, respectively. {\bf d,} Band structures of Mo$_{0.2}$W$_{0.8}$Te$_2$ along momentum space cut as defined in pane (b). We find an inversion between the b$_2$ and b$_3$ bands, giving rise to the Weyl nodes W1 and W2 with  opposite chiral charges $C_w$. {\bf e,} The $k$ space separation between the pair of Weyl nodes W1 and W2 as a function of Mo doping $x$.  {\bf f,} The energy difference between the extrema of the b2 and b3 bands as a function of Mo doping $x$. 
This characterizes the magnitude of the band inversion. {\bf g,} The energy offset between the pair of Weyl nodes W1 and W2 as a function of Mo doping $x$. {\bf h,} A schematic for the distribution of the Weyl nodes within a bulk BZ. All nodes are located on the $k_z=0$ plane.  
}
\label{WP}
\end{figure*}

We now calculate the band structure of pure WTe$_2$ throughout the bulk BZ based on the lattice constants reported in \cite{WT-Structure}. Our results show that pure WTe$_2$ has a continuous energy gap throughout the bulk BZ without any Weyl nodes. The $k$ point that corresponds to the minimal gap is found to be close to the $\Gamma-Y$ (Fig.~\ref{WP}(a)) axis. The minimal gap of WTe$_2$ is $0.9$ meV (Fig.~\ref{WP}(c)). We note that the discrepancy between our results and Ref.~\cite{WT-Weyl} is due to the slightly different values of the lattice constants \cite{WT-Structure, WT-Structure-2}. The lattice constants used in Ref. \cite{WT-Weyl} were at low temperatures \cite{WT-Structure-2}. Thus the results \cite{WT-Weyl} better refelct the groundstate ($T=0$) of WTe$_2$. We used the lattice constants at room temperatures \cite{WT-Structure}, so our results correspond to the state of WTe$_2$ at elevated temperatures. The difference between our results and Ref.~\cite{WT-Weyl} shows from another angle that WTe$_2$ 
is very close to the phase boundary between the Weyl state and the fully gapped state. For many purposes, it is favorable to have the Weyl state in a material robust at elevated (room) temperatures. Here, we use the room temperature lattice constants for all of our calculations at all Mo concentrations. We also note that the very small difference of the lattice constant value does not play a role except for undoped or very lightly doped samples $x\leq2$, where the separation of the Weyl nodes is beyond experimental resolution anyway. 


We propose Mo doped WTe$_2$, Mo$_{x}$W$_{1-x}$Te$_2$, as an experimental feasible platform to realize Weyl state in this compound. We have shown that pure WTe$_2$ is very close to the phase transition boundary. Therefore, the $k$ splitting between the Weyl nodes would be beyond experimental resolution. On the other hand, another very recent paper proposed a Weyl state in pure MoTe$_2$ \cite{MT-Weyl}, but as shown above the existence of the orthorhombic MoTe$_2$ needs to be confirmed. By contrast, we show that the moderately Mo doped WTe$_2$ sample have a number of advantages, making it experimentally feasible. First, pure MoTe$_2$ has many irrelevant band crossing the Fermi level along the $\Gamma-X$ and $\Gamma-Z$ directions, whereas the band structure of moderately Mo-doped system is as clean as pure WTe$_2$ (Figs.~\ref{band}(c-e)). Second, as we will show below, a moderate Mo doping leads to a $k$ space separation of the Weyl nodes that is similarly large as pure MoTe$_2$. Therefore, we propose the Mo-
doped WTe$_2$ as a better platform for studying Weyl physics. Fig.~\ref{WP}(e) shows the evolution of the $k$ space distance between a pair of Weyl nodes as a function of Mo concentration. Our calculation shows that a 2\% Mo doping is sufficient to stabilize the system in the Weyl metal state. Also, the distance between the Weyl nodes increases rapidly at the small doping regime. At a moderate doping $x=20\%$, the $k$ space distance is found to be as large as $0.03 \frac{2\pi}{a}$. As one further increases the doping concentration, the distance seems saturated. The distance is about $0.04 \frac{2\pi}{a}$ at $x=40\%$. The energy difference between the pair of Weyl nodes is shown in Fig.~\ref{WP}(g). In Fig.~\ref{WP}(d) we show the dispersion along the momentum space cut that goes through the direct pair of Weyl nodes as defined in Fig.~\ref{WP}(b). It can be seen clearly that two singly generate bands, b2 and b3, cross each other and form the two Weyl nodes with opposite chiralities. We name the Weyl node at 
lower energy as W1 and the Weyl node at higher energy as W2. Another useful quantity is the energy difference between the extrema of these two bands. This characterizes the magnitude of the band inversion, as shown in Fig.~\ref{WP}(f). It is interesting to note that, in contrast to the $k$ space distance between the Weyl nodes (Fig.~\ref{WP}(e)), the energy difference between the Weyl nodes (Fig.~\ref{WP}(g)) and the band inversion energy (Fig.~\ref{WP}(f)) does not show signs of saturation as one increases the Mo concentration $x$ up to $40\%$. In Fig.~\ref{WP}(h), we show a schematic for the distribution of the Weyl nodes in Mo doped WTe$_2$. We observe a pair of Weyl nodes in each quadrant of the $k_z=0$ plane. Thus in total there are 4 pairs of Weyl nodes on the $k_z=0$ plane. 

\begin{figure*}[t]
\includegraphics[width=17cm]{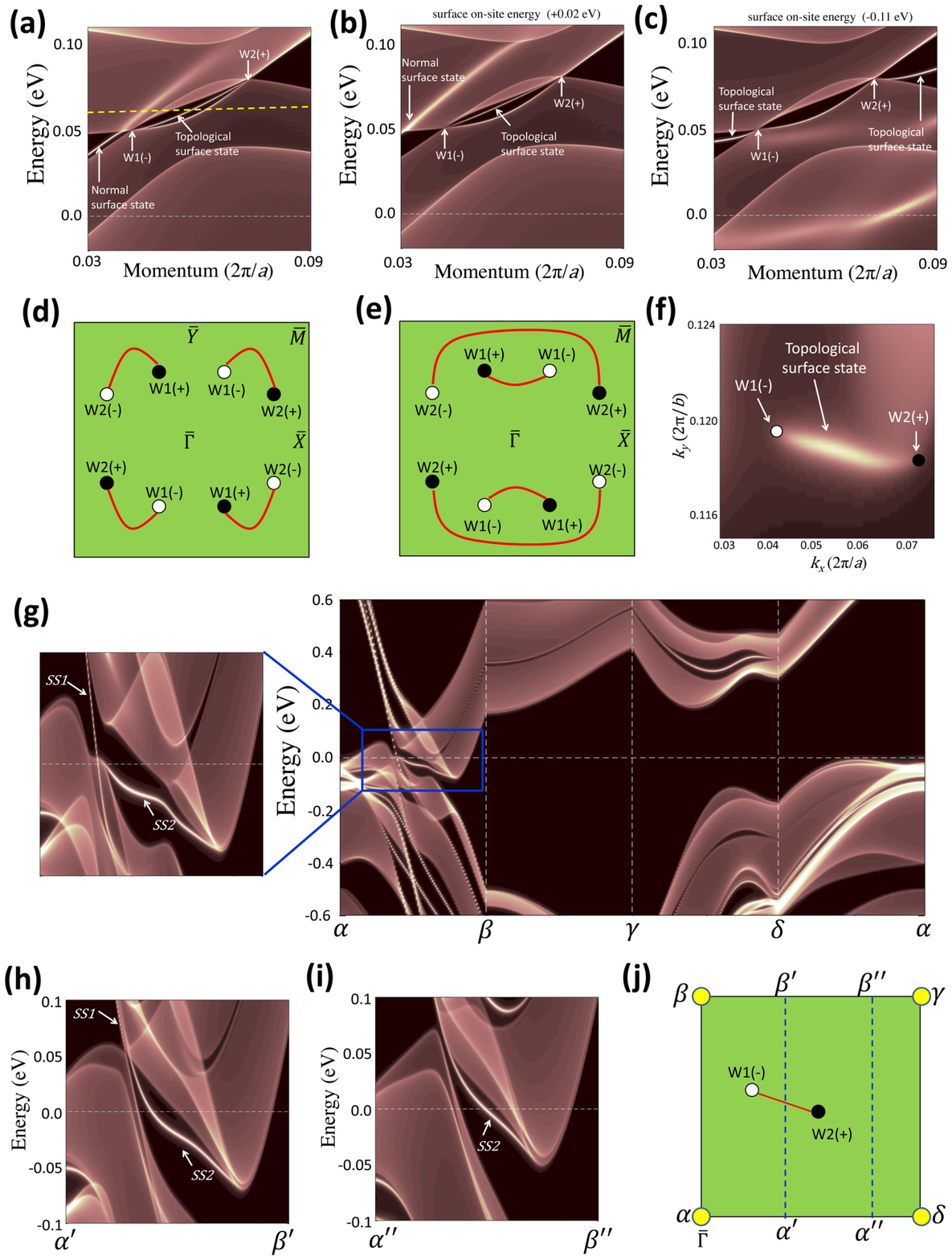}
\caption{ 
{\bf Tunable Fermi arc length and inter-connectivity in Mo$_{x}$W$_{1-x}$Te$_2$.}}
\label{Fig_FSv5}
\end{figure*}

A critical signature of a Weyl semimetal/metal is the existence of Fermi arc surface states. We present calculations of the (001) surface states in Fig.~\ref{Fig_FSv5}. We choose the 20\% Mo-doped system, Mo$_{0.2}$W$_{0.8}$Te$_2$. Figure~\ref{Fig_FSv5}(a) shows the surface energy dispersion along the momentum space cut that goes through the direct pair of Weyl nodes, W1(-) and W2(+), which arises from a single band inversion. Our calculation (Fig.~\ref{Fig_FSv5}(a)) clearly shows the topological Fermi arc surface state, which connects the direct pair of Weyl nodes. The Fermi arc is found to terminates directly onto the projected Weyl nodes. In addition, we also observe a normal surface state, which avoids the Weyl node and merges into the bulk band continuum. Because the W1 and W2 Weyl nodes have different energies, and because W1 is a type II Weyl cone \cite{WT-Weyl}, constant energy maps always have finite Fermi surfaces. Hence, visualizing Fermi arc connectivity in constant energy maps is not 
straightforward. Instead of a constant energy map, it is possible to use a varying-energy $k_x,k_y$ map, i.e. Energy$=E(k_x,k_y)$, so that there are no bulk states on this varying-energy map at all $k_x,k_y$ points except the Weyl nodes. Figure~\ref{Fig_FSv5}(f) shows the calculated surface and bulk electronic structure on such a varying-energy $k_x, k_y$ map in the vicinity of a pair of Weyl nodes. A Fermi arc that connects the pair of Weyl nodes can be clearly seen. We study the effect of surface perturbations. The existence of Weyl nodes and Fermi arcs are guaranteed by the system's topology whereas the details of the surface states can change under surface perturbations. In order to do so, we change the surface on-site potentials of the system. Physically, the surface potentials can be changed by surface deposition or applying an electric field on the surface. Figure~\ref{Fig_FSv5}(b) shows the surface band structure with the surface on-site energy increased by 0.02 eV. We find that the normal surface 
state moves further away from the Weyl nodes whereas the topological Fermi arc does not change significantly. Figure~\ref{Fig_FSv5}(c) shows the surface band structure with the surface on-site energy decreased by 0.11 eV. The normal surface states disappear. The Fermi arc also changes significantly. Instead of directly connecting the two Weyl nodes in Figs.~\ref{Fig_FSv5}(a and b), a surface state stems from each Weyl node and disperses outside the window. We note that the surface states in Fig.~\ref{Fig_FSv5}(c) are still topological and are still arcs because they terminate directly onto the projected Weyl nodes. We illustrate the two types of Fermi arc connectivity in Figs.~\ref{Fig_FSv5}(d and e). Figure~\ref{Fig_FSv5}(d) corresponds to the case in Figs.~\ref{Fig_FSv5}(a and b), where a Fermi arc directly connects the pair of Weyl nodes in a quadrant. Figure~\ref{Fig_FSv5}(e) corresponds to Fig.~\ref{WP}(d). In this case, Fermi arcs connect Weyl nodes in two different quadrants across the $\bar{Y}-\bar{\
Gamma}-\bar{Y}$ line. The nontrivial topology in a Weyl semimetal requires that there must be Fermi arc(s) terminating onto each projected Weyl node with a nonzero projected chiral charge and that the number of Fermi arcs associated with a projected Weyl node must equal its projected chiral charge. On the other hand, the pattern of connectivity can vary depending on details of the surface. The observed different Fermi arc connectivity patterns as a function of surface on-site potential provide an explicit example of both the constraints imposed and the freedoms allowed to the Fermi arc electronic structure by the nontrivial topology in a Weyl semimetal.

We further study the surface states via bulk boundary correspondence. We note that except Figs.~\ref{Fig_FSv5}(b-e), all other figures corresponds to the case without additional changes to the onsite energy. Specifically, we choose a closed loop in ($k_x, k_y$) space as shown in Fig.~\ref{Fig_FSv5}(j). As we mentioned above, the conduction and valence bands only touch at the eight Weyl nodes. Thus as long as the loop chosen does not go through these Weyl nodes, there is a continuous bulk energy gap along the loop. In the bulk BZ, the chosen rectangular loop corresponds to a rectangular pipe along the $k_z$ direction. Then topological band theory requires that the net chiral charge of the Weyl nodes that are enclosed by the pipe equals the Chern number on this manifold, which further equals the net number of chiral edge modes along the loop. For example, the rectangular loop $\alpha-\beta-\gamma-\delta$ in Fig.~\ref{Fig_FSv5}(j) encloses a W1(-) and a W2(+), which leads to a net chiral charge zero. The energy 
dispersion along this rectangular loop is shown in Fig.~\ref{Fig_FSv5}(g). It can be seen that the bands are fully gapped without any surface states along $\beta-\gamma-\delta-\alpha$. Along $\alpha-\beta$, there are two surface states, SS1 and SS2, both of which connect the band gap. Interestingly, we note that these two surface bands are counter-propagating although they seem to have the same sign of Fermi velocity. This is because the continuous energy gap $\alpha-\beta$ is highly ``tilted''. If we ``tilt'' the energy gap back to being horizontal, then it can be clearly seen that the two surface bands are counter-propagating, which means that the net number of chiral edge modes is zero. Similarly, we can choose other loops. For example, we choose another rectangular loop $\alpha'-\beta'-\gamma-\delta$ that encloses only the W2(+) Weyl node. Because there are no surface states along the two horizontal edges and the vertical edge to the right, we only need to study the vertical edge to the left, that is the 
$\alpha'-\beta'$. The enclosed net chiral charge is $+1$, which should equal the net number of chiral edge mode along $\alpha'-\beta'$. The band structure along this line is shown in Fig.~\ref{Fig_FSv5}(h). We see that while the surface band SS1 still connects across the band gap, SS2 starts from and ends at the conduction bands. Therefore, SS1 contributes one net chiral edge mode whereas SS2 contributes zero net chiral edge mode. Hence there is one net chiral edge mode along the this rectangular loop. By the same token, we can choose the loop $\alpha''-\beta''-\gamma-\delta$, which does not enclose any Weyl node. Consistently, as shown in Fig.~\ref{Fig_FSv5}(i), along $\alpha''-\beta''$, SS1 does not appear along this line and SS2 does not connect across the band gap. Hence the net number of chiral edge mode is also zero along this rectangular loop.

we study the constant energy contours of the surface states. We emphasize that (1) there is a significant energy offset between the W1 and W2 Weyl nodes, and that (2) the W1 Weyl cones are the type II Weyl cone \cite{WT-Weyl}, which means that at the Weyl node energy, its constant energy contour consists of an electron and a hole pockets touching at a point, the Weyl node. These two properties are very different from the ideal picture, where all Weyl cones are normal rather than type II and their nodes are all at the same energy. We show below that these two properties make the surface states' constant energy contours quite different from what one would expect naively. Fig.~\ref{schematic}(d) shows the calculated constant energy contour within the top half of the surface BZ at energy $E_1$, which is between the W1 and W2 nodes in energy. We see three bulk pockets. The corresponding schematic is shown in Fig.~\ref{schematic}(a). Specifically, we see a big pocket closer to the $\bar{Y}$ point, which encloses 
two W1 Weyl nodes with opposite chiral charges. We also see two separate small pockets closer to the $\bar{\Gamma}$, each of which encloses a W2 Weyl node. As for the surface states, from Figs.~\ref{schematic}(d and e) we see a surface state band that connects the two small pockets, each of which encloses a W2 Weyl node. This is quite counter-intuitive because we know that the Fermi arc connects the direct pair of Weyl nodes, namely a W1(-) and a W2(+) or vice versa. We show that there is no discrepancy. Specifically, we show that the surface band seen in the constant energy contours is exactly the Fermi arc that connects the W1(-) and the W2(+) Weyl nodes seen in Fig.~\ref{schematic}(c). To do so, we consider the constant energy contours at two different energies, $E_1$ and $E_2$. According to the energy dispersion (Fig.~\ref{schematic}(c)), we see that the big bulk pocket in the constant energy map is electron-like while the two small bulk pockets are hole-like. Thus as we increase the energy from $E_1$ to 
$E_2$, the big pocket should expand whereas the two small pockets should shrink, as shown in Fig.~\ref{schematic}(b). The surface state band keeps connecting the two small pockets as one changes the energy. This evolution is shown by real calculations in Figs.~\ref{schematic}(e and f). The orange line in Fig.~\ref{schematic}(b) connects the W1(-) and W2(+) Weyl nodes. At each energy, the surface state band crosses the orange line at a specific $k$ point. By picking up the crossing points at different energies, we can reconstruct the Fermi arc that connect the W1(-) and W2(+) Weyl nodes shown in Fig.~\ref{schematic}(c). Therefore, from our systematic studies above, we show that the Fermi arc connectivity means the pattern in which the surface state connects the Weyl nodes. This is defined on a varying-energy ($k_x,k_y$) map where the chosen $E(kx,ky)$ map crosses the bulk bands only at the Weyl nodes. If there is no significant energy offset between Weyl nodes and if all Weyl cones are normal rather than type 
II, then the connectivity can also be seen in a constant energy contour. However, in our case here, one needs to be careful with the simplified ideal picture, that is to study the Fermi arc connectivity from the constant energy contour. Because of the energy offset between the Weyl nodes and because of the existence of type II Weyl cones, how surface bands connect different bulk pockets in a constant energy contour does not straightforwardly show the Fermi arc connectivity. 

Finally, we discuss the tunability of the length of the Fermi arcs as a function of Mo concentration $x$ in our Mo$_{x}$W$_{1-x}$Te$_2$ system. (1) The undoped $x=0$ sample is fully gapped according to our calculations (Fig.~\ref{schematic}(g)). (2) A very small Mo concentration ($\sim0.5\%$) will drive the system to the critical point, where the conduction and valence bands just touch each other Fig.~\ref{schematic}(h). The length of the Fermi arc is zero, and hence the system is at the critical point. (3) As one further increases the Mo concentration $x$, the touching point splits into a pair of Weyl nodes with opposite chiralities (Fig.~\ref{schematic}(i)). The Weyl nodes are connected by a Fermi arc. A way to gap the system without breaking any symmetry is to annihilate pairs of Weyl nodes with opposite chiralities. In order to do so, one needs to overcome the momentum space separation between the Weyl nodes to bring them together in $k$ space. For a sample with a given Mo concentration $x$, this may be 
achieved by applying external hydrostatic pressure. Thus the length of the Fermi arc provides a measure of the system's topological strength. Such a tunability is not known in the TaAs class of Weyl system. This highlights the uniqueness of our work. 

\textit{Acknowledgements:} 
T.R.C. and H.T.J. are supported by the National Science Council, Academia Sinica, and National TsingHua University, Taiwan. We also thank NCHC, CINC-NTU, and NCTS, Taiwan for technical support. Work at Princeton University were supported by the Gordon and Betty Moore Foundations EPiQS Initiative through Grant GBMF4547 (Hasan). Work at National University of Singapore were supported by the National Research Foundation, Prime Minister's Office, Singapore under its NRF fellowship (NRF Award No. NRF-NRFF2013-03). The work at Northeastern University was supported by the US Department of Energy (DOE), Office of Science, Basic Energy Sciences grant number DE-FG02-07ER46352, and benefited from Northeastern University's Advanced Scientific Computation Center (ASCC) and the NERSC Supercomputing Center through DOE grant number DE-AC02-05CH11231. We thank B. A. Bernevig, Chen Fang, and Titus Neupert for discussions. 


\addtocounter{figure}{-1}
\begin{figure*}[t!]
\caption{ {\bf a,} Surface and bulk band structure of Mo$_{0.2}$W$_{0.8}$Te$_2$ along the momentum space cut that goes through a direct panel of Weyl nodes, W1(-) and W2(+), as defined in Fig. \ref{WP}(b). A topological Fermi arc surface state connects the Weyl nodes W1 and W2. A normal surface state avoids the Weyl nodes and merges into the bulk band continuum. {\bf b,} Same as panel (a) but with the surface on-site energy increased by 0.02 eV. {\bf c,} Same as (a) but with the surface on-site energy decreased by 0.11 eV. {\bf d,} Schematic of Fermi arc connectivity pattern for panels (a and d). {\bf e,} Schematic of Fermi arc connectivity pattern for panel (c). {\bf f,} Surface band structure on a varying-energy ($k_x,k_y$) map. Specifically, we choose a different energy on each ($k_x,k_y$) point, so that the there are no bulk states at all $k_x,k_y$ except at the locations of the eight Weyl nodes. This is possible because we know that the conduction ad valence bands only cross at the eight Weyl nodes. {\
bf g,} Surface and bulk band structure along the $k-$space trajectory $\alpha-\beta-\gamma-\delta$ defined in panel (j). {\bf h,i,} Surface and bulk band structure along the $k-$space trajectories $\alpha'-\beta'$ and $\alpha'-\beta'$ defined in panel (j). {\bf j,} Schematic of a quadrant of the surface BZ, showing the $k-$space trajectories used in panels (g-i).}
\end{figure*}

\begin{figure*}[t]
\includegraphics[width=17cm]{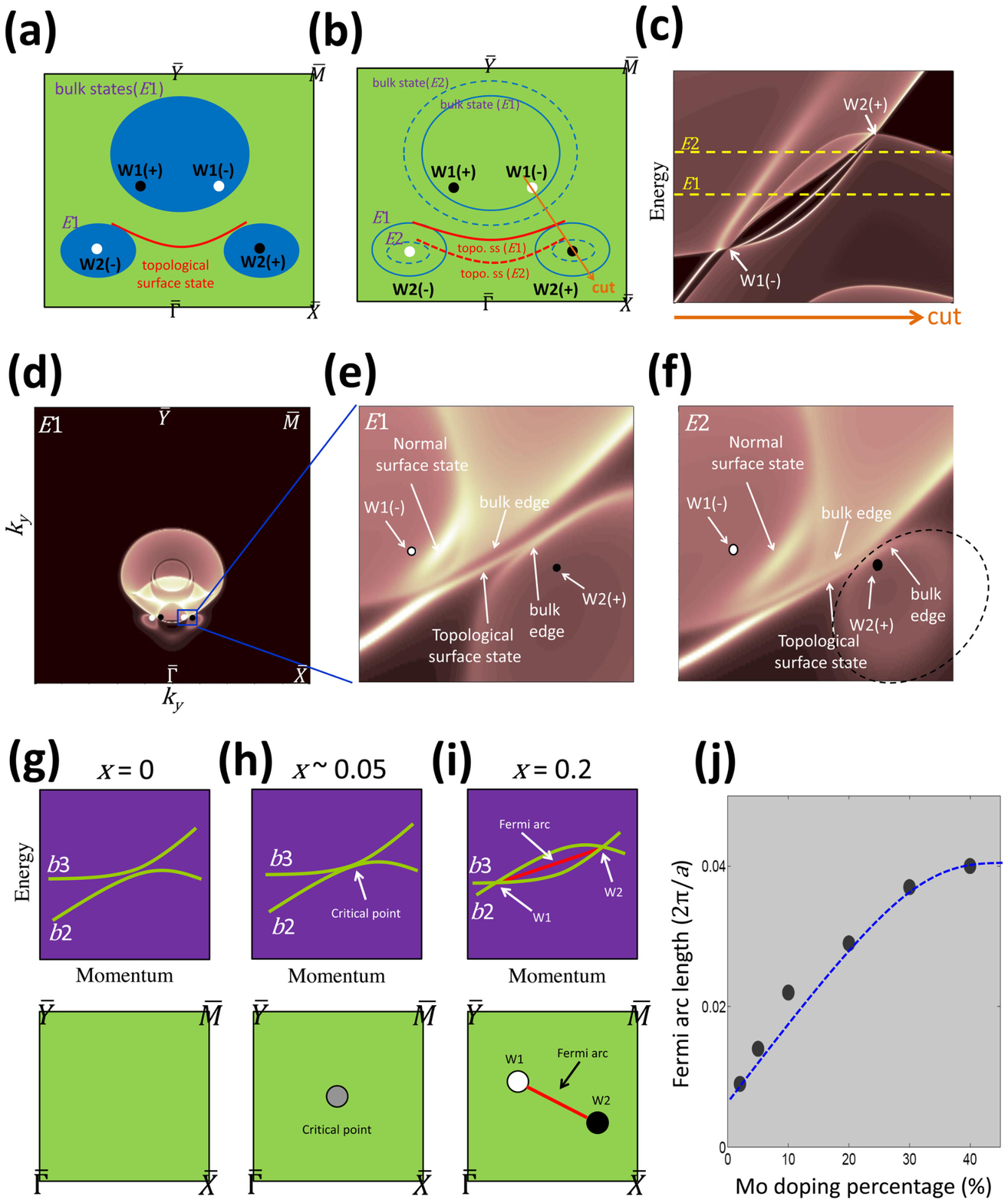}
\caption{ 
{\bf Iso-energy contour interconnectivity on the surface of Mo$_{x}$W$_{1-x}$Te$_2$.} {\bf a,} Schematic illustration of the surface and bulk electronic structure on a constant energy $E_1$. The shaded areas represent the projected bulk bands whereas the red line show the surface states.}
\label{schematic}
\end{figure*}
\addtocounter{figure}{-1}
\begin{figure*}[t!]
\caption{{\bf b,} Schematic illustration of the surface and bulk electronic structure on two energies $E_1$ and $E_2$. The band structure on $E_1$ ($E_2$) are shown by the solid (dotted) lines. The energies $E_1$ and $E_2$ are defined in panel (c). {\bf c,} Band structure along a $k$ space cut that goes through the direct pair of Weyl nodes, W1(-) and W2(+). The dotted lines denote the energies $E_1$ and $E_2$ with respect to the W1 and W2 Weyl nodes. {\bf d,} Calculated surface and bulk electronic structure on at the energy $E_1$ over the top half of the surface BZ. {\bf e,} A Zoomed-in view of panel (d) for the area highlighted by the blue box. {\bf f,} The same as panel (e) but at the energy $E_2$. {\bf g-i,} Schematic band diagram to show the evolution of the Mo$_{x}$W$_{1-x}$Te$_2$ system as a function of Mo concentration $x$. {\bf j,} The length of the Fermi arc as a function of Mo concentration $x$. The arc length equals the $k$ space separation of the Weyl nodes.}
\end{figure*}
\end{document}